\documentclass[confrence]{IEEEtran}
\IEEEoverridecommandlockouts
\IEEEpubid{\makebox[\columnwidth]{978-1-6654-1216-2/22/\$31.00~\copyright~2022 IEEE}
	\hspace{\columnsep}\makebox[\columnwidth]{ }}
\usepackage{float}
\usepackage{cite}
\usepackage{amsmath,amssymb,amsfonts}
\usepackage{algorithmic}
\usepackage{graphicx}
\usepackage{textcomp}
\usepackage{xcolor}
\usepackage{tabularx}
\usepackage{algorithmic}
\usepackage[linesnumbered,ruled,vlined]{algorithm2e}
\SetKwInput{KwInput}{Input}
\SetKwInput{KwOutput}{Output}
\def\BibTeX{{\rm B\kern-.05em{\sc i\kern-.025em b}\kern-.08em
		T\kern-.1667em\lower.7ex\hbox{E}\kern-.125emX}}
	 	
\begin{document}

	\title{Deep Curriculum Learning for PolSAR Image Classification\\}
		%{\footnotesize \textsuperscript{*}Note: Sub-titles are not captured in 
		%Xplore and should not be used}

	\author{\IEEEauthorblockN{Hamidreza Mousavi, Maryam Imani, Hassan Ghassemian\\}
		\IEEEauthorblockA{\textit{Image processing and Information Analysis Lab } \\
			\textit{Faculty of Electrical and Computer Engineering, Tehran, Iran}\\
			\{h.mousavi, maryam.imani, ghassemi\}@modares.ac.ir}}

\maketitle

	%%%%%%%%%%%%%%%%%% Abstract %%%%%%%%%%%%%%%%%%%%%%%%%%%%%%%%%%%%%%%
	\begin{abstract}
		Following the great success of curriculum learning in the area of machine learning, a novel deep curriculum learning method proposed in this paper, entitled  DCL, particularly for the classification of fully polarimetric synthetic aperture radar (PolSAR) data. This method utilizes the entropy-alpha target decomposition method to estimate the degree of complexity of each PolSAR image patch before applying it to the convolutional neural network (CNN). Also, an accumulative mini-batch pacing function is used to introduce more difficult patches to CNN.
		Experiments on the widely used data set of AIRSAR Flevoland reveal that the proposed curriculum learning method can not only increase classification accuracy but also lead to faster training convergence.
	\end{abstract}
	
	\begin{IEEEkeywords}
		Polarimetric SAR image classification, Curriculum learning, Deep learning, Convloutional neural network.
	\end{IEEEkeywords}

	%%%%%%%%%%%%%%%%%% Introduction %%%%%%%%%%%%%%%%%%%%%%%%%%%%%%%%%%%%%%%
	\section{Introduction}
	Deep neural networks have been extensively employed in the area of PolSAR image classification. Specially convolutional neural networks (CNNs) have proved to be a useful automatic feature extraction tool for PolSAR image classification in recent years\cite{b1}. The main focus in this area of research is on developing practical CNN architectures in terms of classification accuracy and model complexity. For instance, Zhang \textit{et al.}\cite{b2} extended a real-valued CNN (RV-CNN) in a complex domain to exploit the potential of phase information in PolSAR data and their proposed complex-valued CNN (CV-CNN) could remarkably increase the classification accuracy over RV-CNN. The first architecture utilizing 3D convolution and 3D pooling layers for PolSAR image classification was introduced by Zhang \textit{et al.}\cite{b3} and achieved a classification accuracy better than conventional 2D CNNs. Lightweight 3D convolutions introduced by Dong \textit{et al.}\cite{b4} to reduce the redundancy of 3D convolutions and take advantage of their excellent ability in spatial-polarimetry feature extraction. Considerable attention has been focused on the CNN architecture evolution over the last few years for increasing the feature extraction ability. In contrast, to reduce the number of model parameters, prevent from the overfitting, and even to eliminate the fully connected layers, less attention has been devoted to improving the training process. Since neural network architectures are inspired by the human brain, it seems reasonable that studying the learning process should also be inspired by the biological processes of the human brain. Starting from easy tasks before progressing to high and complicated levels generally provides better learning results for Humans. Using a similar approach for training a machine learning model, we can obtain great benefits including faster convergence speed of the training process and better accuracy. This strategy has been formalized by Bengio \textit{et al.}\cite{b5} and called \textit{curriculum learning} in the context of machine learning. Curriculum learning approaches have been successfully applied in various fields of machine learning\cite{b6}. Each curriculum learning strategy needs a proper complexity measurement criterion of training samples and a pacing function for progressively introducing more complex samples to the machine learning model\cite{b5,b6}.
	To our knowledge, there isn't a criterion to measure the difficulty of PolSAR patches before applying them to a deep learning model. This motivates us to formalize this idea.\\
	PolSAR data analysis is rich in accurate pixel information extraction thanks to the publication of a large number of proposed methods in the target decomposition theory. The Cloude-Pottier target decomposition method proposed parameters such as entropy (H), mean alpha angle ($\bar\alpha$), and anisotropy (AA), which provide the physical interpretation of the scattering mechanisms of objects. The most attractive and practical application of the H/$\bar\alpha $ method is an unsupervised classification of PolSAR data by subdividing the H/$\bar\alpha $ plane into nine zones correspond to various physical scattering characteristics. In this study, we use the H/$\bar\alpha $ plane to introduce a novel patch complexity criterion (PaCC) for PolSAR patches.
	The basic idea is that each patch is considered more difficult for neural network to learn if it consists of pixels with a more complex scattering process. Therefore, two criteria are needed to first measure the complexity of each pixel in the patch and then determine the complexity of the patch with the obtained values for pixels. Two parameters of entropy H and mean alpha angle $\bar\alpha$, which yield direct information about the scattering mechanism are used to measure the scattering complexity of pixels and a criterion named pixel complexity criterion (PiCC) is defined. Finally, by calculating the average of the obtained PiCC values, the PaCC parameter is determined.
	The rest of this paper is structured as follows. Section II presents the methodology. Experimental results are shown in Section III, which illustrate the effectiveness of the proposed method. Finally, we draw some conclusions in Section IV.
	\section{METHODOLOGY}\label{meth}
	
	\subsection{Deep learning}\label{CNN}
	Deep learning (DL) attempts to develop suitable models by mimicking the structure and functioning of the human brain in processing data. From the viewpoint of automation, the evolution of DL has transformed the handcrafted design of features in machine learning to facilitate automatic feature extraction from high-dimensional raw data. This remarkable ability of high-dimensional data processing and automatic feature extraction has made breakthroughs in the pattern recognition field. Most recently, The huge success of DL in the computer vision area has motivated the remote-sensing community to this technology and DL algorithms bring huge advancements in many remote sensing image interpretation tasks including Image fusion\cite{b11}, scene classification \cite{b17}, object detection\cite{b12,b16}, surface terrian classification\cite{b18}, and other tasks.
	Convolutional neural networks (CNN) is a specific type of DL model which is particularly useful for recognizing visual patterns in images. It is inspired by the structure of the human and animal visual system and designed to automatically extract low, mid and high-level features from multi-dimensional data. Because of this characteristic, it is applicable for processing multi-channel remote sensing images.
	
	CNN has proven to have an exemplary performance to fill the gaps in conventional machine learning algorithms.\\
	Inspired by the lightweight operations for CNNs in the PolSAR image classification field, in this paper, pseudo-3D(P3D) CNN  introduced by \cite{b4} is used in the experiments to investigate the performance of the proposed curriculum learning algorithm. Lightweight P3D CNN was introduced to not only perform feature learning in both the spatial and polarimetric dimensions but also reduce network parameters and high computational costs of 3D CNNs\cite{b4,b15}.
	
	\subsection{Curriculum Learning}\label{Cur learning}  
	Curriculum Learning (CL) can be considered as an optimization technique, which can improve optimization performance by discovering better relative minima of a non-convex cost function\cite{b5,b7}. It is inspired based on imitating the learning behaviour of humans when trying to learn concepts. They can master a task when try to learn easy parts of a task and gradually increase 
	complexity level. Using a similar approach for training a machine learning model, we can obtain great benefits including faster convergence speed of the training process and better accuracy. \\
	Since that samples are considered in random order during training of machine learning model, the hard or noisy samples, especially in the initial stage of the training process may result in over-fitting the train set and leading to poor generalization. This learning strategy has been empirically proved to be robust to noisy samples and be effective in avoiding getting stuck in poor local minima\cite{b14}. PolSAR images also are heavily affected by speckle noise. This noise diminishes the visual quality of PolSAR images for interpretation and limits image processing algorithms during information extraction from PolSAR data. Hence, CL can effectively help machine learning algorithms to suppress the effect of noisy PolSAR samples during the training process and achieve more reliable results. However, the implementation of CL in machine learning is subject to finding a proper computational complexity measurement criterion of training samples and a pacing function for progressively introducing more complex samples to the machine learning model. This is the principal motivation behind the design and implementation of our CL framework to improve the classification performance of SAR polarimetry data.\\
	The complexity of the scattering process complicates the direct analysis of coherency matrix T. Hence, the ultimate purpose of the incoherent decompositions tools is to simplify the interpretation of the scattering process by expanding the second-order matrices into a weighted addition of basis matrices describing scattering responses of simpler objects.
	Two important average parameters (the entropy H and mean  angle $\bar\alpha$) were introduced by Cloude and Pottier\cite{b7} using the eigenvalue/eigenvector analysis of T. Entropy H is defined as a function of the eigenvalues:
	\\
	\begin{equation*}\tag{1}
		\begin{aligned}
			H&=-\sum^{3}_{i=1}p_i\log_{3}(p_i)     & , && p_i &=\dfrac{\lambda_i}{\sum^{3}_{j=1} \lambda_{j}}\\
		\end{aligned}
	\end{equation*}
	\\
	Entropy H can go above 0 and below 1. H = 0 corresponds to only one nonzero eigenvalue in T. It physically indicates that there is only a single scattering mechanism caused by the surface scattering. When H equals to 1, T has three equal nonzero eigenvalues, which statically means three scattering mechanisms occur simultaneously. From a physical standpoint, this shows that the behaviour of scatter is completely random and this mechanism occurs over the forest and vegetated surfaces. As H increases from 0 to 1, the randomness of the scattering mechanism increases, which means the ability to distinguish between scattering mechanisms decreases\cite{b8,b9}. In short, the the complexity of every pixel in the PolSAR image is proportional to entropy H.\\
	Mean alpha angle $\bar\alpha$ is obtained from parameterizing of the eigenvectors of T versus some of the angle parameters which $\alpha$ is one of them and given by 
	\\
	\begin{equation*}\tag{2}
		\bar\alpha = \sum^{3}_{i=1}p_i{\alpha_i}
	\end{equation*}
	\\
	The mean alpha angle $\bar\alpha$ indicates the type of dominant scattering mechanism and lies within the range of $0^\circ$ to $90^\circ$. $\bar\alpha$ = $0^\circ$  corresponds to the surface scattering caused by water body or bare soil, $\bar\alpha$ = $45^\circ$ to the volumetric scattering arisen from woodlands, and $\bar\alpha$ = $90^\circ$ to double-bounce scattering as a result of interaction between the incident wave and street/building classes in urban areas. By continuously increasing  $\bar\alpha$ from 0 to 90, the scattering mechanism changes from isotropic simple scattering($\bar\alpha$ = 0) to random volume scattering ($\bar\alpha$ = 60) and finally  $\bar\alpha$ = 90, is related to double bounce scattering\cite{b7,b8}.
	For the measurement of complexity based on the alpha angle, if we consider complexity range from 0 to 1, by increasing $\bar\alpha$ from 0 to 60, complexity increases to 1 and by changing alpha from 60 to 90, the complexity decreases from the maximum value to the median one, which equals to 0.5. Based on the above analysis, the PaCC parameter is proposed to measure the complexity of each PolSAR image patch by averaging the complexity of whole pixels exist in that patch. The complexity of each pixel is calculated by using Euclidean distance in the two-dimensional (2D) H/$\bar\alpha$ plane. The details of the patch-ranking algorithm are shown in Algorithm \ref{algorithm1} and the details of the method are expressed as follows.\\
	First, we introduce a metric, called pixel complexity criterion (PiCC) to measure the complexity of every pixel in a patch using Euclidean distance as follows:
	\\
	\begin{equation*}\tag{3}
		\begin{aligned}
			PiCC(\textbf{v}^{(i)}) = &\sqrt{H_i^2 + (1 - |\dfrac{\bar\alpha_i - 60}{60}|)^2}  \quad i= 1,2, ... , m_1m_2 \\
		\end{aligned}
	\end{equation*}
	Where $ \textbf{v}^{(i)}$ is i-th pixel in a PolSAR patch, which defined as a vector as shown in Equation (4) : \\
	\begin{equation*}\tag{4}
		\begin{aligned}
			\textbf{v}=\{t_{11},t_{22},t_{33},\Re{\{t_{12}\}},\Im{\{t_{12}\}},\Re{\{t_{13}\}},\Im{\{t_{13}\}},&\\\Re{\{t_{23}\}},\Im{\{t_{23}\}}\}
		\end{aligned}
	\end{equation*}
	\\
	In equation (3), $H_i$ and $\bar\alpha_i$ are the entropy and mean alpha angle of pixel  $\textbf{v}^{(i)}$, which exist in a PolSAR image patch with $m_{1}  \times  m_{2}  \times9$ size.
	Finally, by averaging on all the calculated PiCCs for the patch, the patch complexity criterion (PaCC) for arbitrary PolSAR image patch $\emph{\textbf{X}} \in \mathbb{R} ^{m_1\times m_2 \times 9}$ is defined by the following formula:
	\begin{multline}\tag{5}
		PaCC(\emph{\textbf{X}}) = \dfrac{1}{m_1m_2}\sum_{i = 1}^{m_1m_2}PiCC(\textbf{v}^{(i)})\\
		= \dfrac{1}{m_1m_2}\sum_{i = 1}^{m_1m_2}\sqrt{H_i^2 + (1 - |\dfrac{\bar\alpha_i - 60}{60}|)^2}        
	\end{multline}
	\\
	Any patch with a higher PaCC value is considered a more difficult patch for training. Therefore, after calculating the PaCC value for all the patches in the training set $\mathbb{L}$, it is necessary to arrange the PaCC values in ascending order. Finally, the training samples are reordered to form the sorted training set $\mathbb{L}^\dagger$.
	The \textit{argsort} operator in the eighth line of Algorithm 1 returns indices of the training samples after sorting PaCC values of them in an ascending order.
	
	Another key part of any curriculum learning method is task scheduling, which specifies how to gradually increase the complexity of tasks and update the training process. Several scheduling strategies have been proposed, where batching, sampling and weighing are the most extensively employed ones\cite{b6}. In this study, we introduce a new batching strategy, called batch accumulation scheduler function. 
	The main idea is that in each iteration after random sampling, we sort the training samples from easy to hard. Then, the sorted training set is split into n same size batches.  The fine-tuning process starts on the easiest batch (k = 1), the second batch is concatenated with the previous one and applied to the deep learning model for updating (k = 2). This process continues until all mini-batches are concatenated together. Eventually, the deep learning model is updated with all training samples existing in the training set (k = n). In step k for gradient updating, the sliced batch from the sorted training set $\mathbb{L}^\dagger$ is sliced as follows:  
	
	\begin{equation*}\tag{6}
		\mathbb{B}_k = \{(\emph{\textbf{X}}_{i},y_{i})\}_{i =j_1}^{j_m}, m = k\dfrac{N}{n},k = 1,2,...,n
	\end{equation*}
	\\
	where $ \mathbb{B}_k $ and m are sliced batch and batch size at k-th step, respectively.
	This pacing function is also inspired by curricula taught in education systems around the world that usually enables learning concepts in progressively increasing levels of complexity while considering and reviewing previously learned concepts.
	
	\begin{algorithm}[!t]\label{algorithm1}
		\KwIn{$\mathbb{L} = \{(\emph{\textbf{X}}_{i},y_{i})\}_{i =1}^{N}$: PolSAR image patch training set.}
		\KwOut{$\mathbb{L}^\dagger = \{(\emph{\textbf{X}}_{j},y_{j})\}_{i =j_1}^{j_N}$: Sorted PolSAR image patch training set.}
		\small
		\renewcommand{\arraystretch}{0.5}{
			$ \textbf{c}_i = 0  $, $ i \in \{1,2,3,...,N \} $ 	 \\
			\For{i $\leftarrow$ 1 to N} 
			{ $ \textbf{d}_k = 0  $, $ k \in \{1,2,3,...,m_1m_2 \} $ \\
				\For{k $\leftarrow$  1 to $m_1m_2$  }
				{
					Calculate $ H_k $ and $\bar\alpha_k$ by formulas 1 and 2 using $\textbf{v}^{(k)}$ as input \\		 
					$ \textbf{d}_k   \leftarrow  \sqrt{H_k^2 + (1 - |\dfrac{\bar\alpha_k - 60}{60}|)^2} $
					
				}
				$ \textbf{c}_i \leftarrow \dfrac{1}{m_1m_2} \sum_{k = 1}^{m_1m_2} \textbf{d}_k $
			}
			$ \textbf{j} = \underset{i}{\mathrm{arg\,sort}}  \textbf{ c}_i$

			\Return{$\mathbb{L}^\dagger = \{(\emph{\textbf{X}}_{i},y_{i})\}_{i =j_1}^{j_N}$}

			\caption{Patch-ranking by $H/\bar\alpha$ decomposition}}
		
	\end{algorithm}

 \subsection{Deep Curriculum Learning (DCL)}\label{Deep_Cur_learning}
 
 Algorithm \ref{algorithm2} presents the pseudo-code of the DCL method. First, the weights of P3D CNN initialized by random values. In the iterative phase, $ N_t $ samples randomly chose from all labeled training set $ \mathbb{U} $ and these samples are added to the training set $ \mathbb{L} $. Training samples in the training set sorted from easy to hard based on Algorithm 1. Then training set divided into n batches and we fine-tune P3D CNN by incrementally accumulating batches.

\SetKwInput{kwInit}{Initalization}    	
\begin{algorithm}[!t]
		\KwIn{\\
			$\mathbb{U}$: All labeled patch set.\\
		
			$ N_{t} $: The number of patches to be selected in each stage of DCL , where t = 0,1,..., T, T is number of stages.\\
			$ n $: Training set splitting parameter\\
			\KwOut{Trained P3D CNN model.}
			\kwInit{\\
		    Initilize weights of P3D CNN by random values.\\
		    
			$ \mathbb{S}_0 $ = \{\}. \\
			$\mathbb{L}$ = \{\}.}\\
		}
		
		\small
		\renewcommand{\arraystretch}{0.5}{
			\For{t $\leftarrow$ 0 to T}
				{
					Select $ N_{t} $ patches randomly from $\mathbb{U}$ and build $ \mathbb{S}_{t} $.\\
					
					Remove Selected patched from $  \mathbb{U}$, i.e, $  \mathbb{U} \leftarrow \mathbb{U}\setminus\mathbb{S}_{t} $.\\
					
					$  \mathbb{L} \leftarrow \mathbb{L}\cup\mathbb{S}_{t} $.\\ 
					
					Sort patches in $ \mathbb{L} $ using Algorithm 1 to obtain sorted patch set $\mathbb{L}^\dagger$.	\\
					
					\For{k $\leftarrow$ 1 to n}
					{
						Slice Batch $\mathbb{B}_k$ by Eq. (6).\\
						P3D CNN traning ($ c = 1 $) or fine-tuning($ c > 1 $) using batch $\mathbb{B}_k$.
					} 
			
		}
			\Return{Trained P3D CNN model.}
			\caption{Deep Curriculum learning (DCL) for PolSAR Image Classification\label{algorithm2}}}
\end{algorithm}
	\section{EXPERIMENTAL RESULTS}\label{EXP}
	In this section, to analyze the performance of the algorithms mentioned in this paper, we conduct experiments on the publicly available PolSAR data set, which is collected by the AIRSAR sensor over the Flevoland area. The overall accuracy (OA) and runtime are also used to evaluation of classification performance.
	\subsection{Experimental data and parameter setting}\label{EXP_data}
	The AIRSAR data is a four-look full polarized image acquired by NASA/JPL using L- band PolSAR sensor was mounted on a NASA DC-8 aircraft on 16 August 1989. It reflects an agricultural area of Flevoland in the Netherlands and it is widely used as a benchmark data for PolSAR image classification. The size of this image is 1024 × 750 pixels and contains 15 categories, which are building, water, forest, bare soil, lucerne, beet, potatoes, grass, rapeseed, barley, three different types wheat, steam beans, peas. The RGB image formed by the Pauli decomposition and the corresponding ground truth image, containing a total of 157296 samples and legend of this data set are shown in Fig. 1. Each colour denotes one category and white region stands for unknown land covers.
	multi-look full polarized image
	\begin{figure}[!t]
		\centerline{\includegraphics[width=0.5\textwidth]{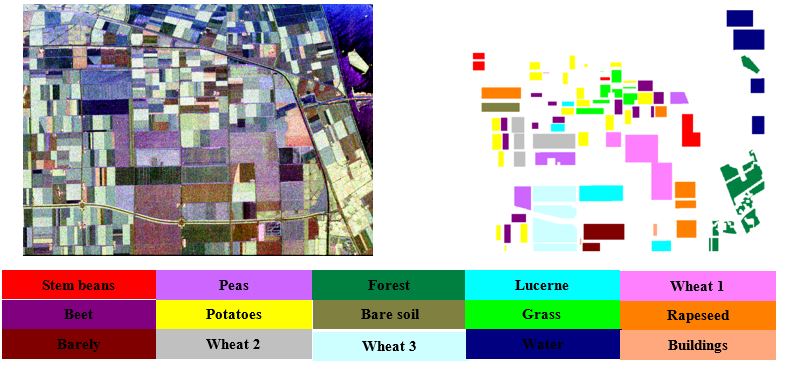}}
		\caption{AIRSAR Floevoland data set adopted in the experiments.}
		\label{fig1}
	\end{figure}
	We set the experimental settings in this work as follows: The lightweight P3D CNN is implemented by Keras with the TensorFlow backend. During the CL cycle, each data set is randomly divided into three sets.  We randomly select 100 labeled samples in each iteration ($ N_t=100 $). To evaluate a model fit while tuning network hyper-parameters, 10000 samples are selected as a validation set and the remaining samples are used for further random sampling. We conduct 30 iterations of network training for the experiments. In each iteration, we join 100 randomly selected training samples to the original training samples to fine-tune the network using our proposed curriculum learning strategy. Therefore, at the end of the training process, a total of 3000 training samples are used. In the proposed curriculum, two hyper-parameters, including splitting parameter n, and fine-tuning iterations need to be chosen. The mentioned two hyper-parameters experimentally are set to 25 and 5, respectively.\\
	We compare our curriculum setting, with a no-curriculum setting in which samples are not ordered by their complexity or applied to P3D CNN by accumulative batching. In other words, during no-curriculum experiment, after each random sampling and before applying to P3D CNN, training samples are shuffled randomly and the training process done by a few epochs with a fixed-batch size. These two hyper-parameters, which are related to no-curriculum setting are set to 25 and 10, respectively. 
    To assure the reliability of experimental outcomes, we do ten independent tests over the data set and the average results are reported.

	\subsection{Classification Results}\label{EXP_data}
	To completely study the influence of curriculum learning on the training process, two groups of experiments are conducted, which are training P3D CNN using curriculum and no-curriculum learning. We have monitored the values of OA during training according to the number of training samples as shown in Fig. 2. The main observations from this figure are as follows:\\
    (a) The OA values with the curriculum, except in a few initial steps are higher than without curriculum. This reflects that curriculum learning provides better generalization to the unseen examples.\\
    (b) With attention to the zoomed area in Fig. 2, the OA value curve for a 
    curriculum learning experiment is smoother than the no-curriculum experiment, which shows that the output results of curriculum-based learning are more reliable for PolSAR data classification. \\
    The obtained OA and running time are given in Table I. From Table I, It is shown that the proposed DCL has higher classification accuracy and lower runtime, which illustrates the strength of the curriculum learning approach and its positive impact on promoting the CNNs.
	\begin{table}[!t]
		\caption{OA VALUES (\%) AND RUN TIMES (SEC) WITH CURRICULUM AND NO-CURRICULUM STRATEGY ON FLEVOLAND AREA DATA SET.}
		\begin{center}
			\begin{tabular}{| c | c | c |}
				\hline
				Method & OA (\%) & run time (sec)
				\\
				\hline
				curriculum   & \textbf{97.64}  &   \textbf{623}
				\\
				\hline
				no-curriculum  & 96.42 &   818 
				\\
				\hline
			\end{tabular}
		\end{center}
	\end{table}
	Finally, for comparative purposes, Fig. 3 (a) and (b) show the classification maps by curriculum learning and no curriculum learning, respectively.  It can be observed that the result obtained by curriculum learning is remarkable, which is better than that obtained by no curriculum learning.
	\begin{figure}[!t]
		\centerline{\includegraphics[width=0.5\textwidth]{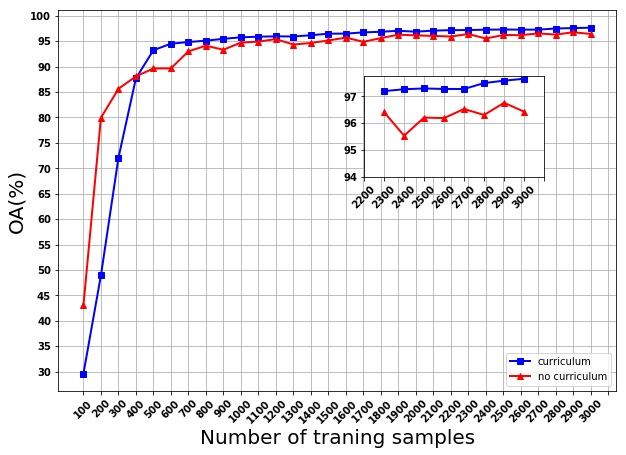}}
		\caption{The obtained overall accuracies (OAs, averaging from 10
			independent runs) using two different strategies for the AIRSAR Flevoland
			data set. In each step, the original training samples are joined with 100 randomly selected training samples. Therefore, at the end of the training process, a total of 3000 training samples are used.}
		\label{fig2}
	\end{figure}

	\section{CONCLUSION}
	In this paper, a novel deep curriculum learning method called DCL has been designed for PolSAR image classification. Besides, a patch complexity criterion, which is called PaCC, is introduced to estimate the learning difficulty of the PolSAR patches before applying them to a CNN.
	The experiments are conducted on benchmark AIRSAR Floevoland data set and the experimental results show that the DCL is fast and can achieve better accuracy than the methods which consider samples in a random order during training.
	
	\begin{figure}[!t]
		\centerline{\includegraphics[width=0.5\textwidth]{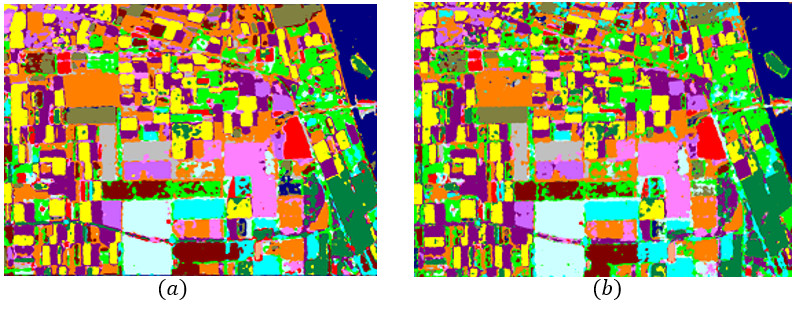}}
		\caption{Classification results of DCL on AIRSAR Floevoland data set. (a) Result of whole classification map using curriculum learning. (b) Result of whole classification map using no curriculum learning.}
		\label{fig3}
	\end{figure}

\end{document}